\documentclass[12pt]{article}
\usepackage{graphicx}
\usepackage[utf8]{inputenc}
\usepackage[english
]{babel}
\def\baselinestretch{1.5}
\begin{document}
\begin{center}
\bf{Probability representation of quantum observable and quantum states}\\
\end{center}
\bigskip

\begin{center} {\bf V. N. Chernega$^1$, O. V. Man'ko$^{1,2}$, V. I. Man'ko$^{1,3}$}
\end{center}

\medskip

\begin{center}
$^1$ - {\it Lebedev Physical Institute, Russian Academy of Sciences\\
Leninskii Prospect 53, Moscow 119991, Russia}\\
$^2$ - {\it Bauman Moscow State Technical University\\
The 2nd Baumanskaya Str. 5, Moscow 105005, Russia}\\
$^3$ - {\it Moscow Institute of Physics and Technology (State University)\\
Institutskii per. 9, Dolgoprudnyi, Moscow Region 141700, Russia}\\
Corresponding author e-mail: manko@sci.lebedev.ru
\end{center}

\section*{Abstract}
We introduce the probability distributions describing quantum
observables in conventional quantum mechanics and clarify their
relations to the tomographic probability distributions describing
quantum states. We derive the evolution equation for quantum
observables (Heisenberg equation) in the pro\-bability
representation and give examples of the spin-1/2 (qubit) states and
the spin observables. We present quantum channels for qubits in the
probability representation.

\medskip

\noindent{\bf Keywords:} quantum suprematism, probability
representation, quantum observables, qubit states, Heisenberg
evolution equation.

\section{Introduction}  
In convectional quantum mechanics, quantum states are identified
either with wave functions (pure states)~\cite{Schr26} or with
density matrices (mixed states)~\cite{Landau,vonNeumann}. The
observables like positions or momenta as well as spin variables are
associated with Hermitian
operators~\cite{Diracbook,Marmobooks1,Marmobooks2} acting in Hilbert
spaces. Other formulations of the quantum-system states associating
with the states the functions on the phase space like the Wigner
function~\cite{Wig32}, the Husimi--Kano
function~\cite{Husimi40,Kano56}, and the Glauber--Sudarshan
function~\cite{Glauber63,Sudarshan63} have been developed to obtain
the formulation of quantum states more similar to the formulation of
the states in classical statistical mechanics.

Recently, the tomographic probability representation of quantum
states was suggested~\cite{Mancini96}; in this representation, the
quantum states are identified with fair probability distributions
connected with density matrices in its phase-space representations
by integral transforms; e.g., the Radon transform~\cite{Radon1917}
of the Wigner function provides the optical
tomogram~\cite{BerBer,VogRis}, which is a standard probability
distribution of continuous homodyne quadrature of photon depending
on an extra parameter called the local oscillator phase, which can
be measured~\cite{RaymerPRL93}.

The probability distributions determining the spin states were
considered
in~\cite{DodPLA,OlgaJetp,Bregence,PainiDariano,Weigert,Wiegert2,22a,22b},
and the tomographic probability representation of quantum states was
studied
in~\cite{Ibort2009,OlgaMarmo2001,SudarshanPascazio1,SudarshanPascazio2,SudarshanPascazio3,my1,my2,my3,MankoPhysScr2015150,RitaNuovoCimento}.

The tomographic probabilities identified with quantum states can be
associated with density operators, in view of the formalism of
star-product quantization~\cite{OlgaMarmo2002,Krakov,my4,my5,my6}
analogous to the procedure where the phase-space quasidistributions
of quantum states, like the Wigner function, are presented within
the star-product framework in~\cite{stratonovich} (see also recent
reviews~\cite{Patrizia,Fedele}). On the other hand, quantum
observables associated with Hermitian operators are presented within
the star-product framework by symbols of the operators, which are
some functions on the phase space, say, in the Wigner--Weyl
representation or the functions of discrete variables in the
spin-tomographic description of qudit states.

The aim of this work is to extend the probability representation of
quantum states to describe the quantum observables in conventional
quantum mechanics by fair probability distributions depending on
extra parameters. Formally, we address the problem of constructing
the invertible map of Hermitian matrices (not only nonnegative
trace-class ones) onto sets of probability distributions depending
on random variables and extra parameters. We construct such
probability representation of quantum observables for systems with
finite-dimensional Hilbert spaces of states which are spin-1/2
systems or systems of qubits.

\section{Probability Representation for Qubit Observables}  
As was shown in~\cite{Marmo,Malevich}, the arbitrary qubit density
2$\times$2 matrix $\rho$ can be presented in the form
\begin{equation} \label{A}
\rho=\left(\begin{array}{cc}
  p_3&p^\ast-\gamma^\ast\\
  p-\gamma & 1-p_3\\
\end{array}\right),\qquad
  \gamma=\frac{1+i}{2}\,,\qquad p=p_1+ip_2,\qquad p_3^\ast=p_3,
\end{equation}
where $0\leq p_1,\,p_2,\,p_3\leq 1$ are the probabilities to have
$m=+1/2$ spin projections on directions $x,$ $y,$ and $z$,
respectively. The three probabilities must satisfy the inequality
\begin{equation} \label{eq.3}
\left(p_1-{1}/{2}\right)^2+\left(p_2-{1}/{2}\right)^2+\left(p_3-{1}/{2}\right)^2\leq {1}/{4}.
\end{equation}
In this section, we demonstrate that an arbitrary spin-1/2
observable can be described by probabilities $0\leq p_1(a)$,
$p_2(a)$, $p_3(a)$, $p_1(b)$, $p_2(b)$, and $p_3(b)\leq 1$, where
$a$ are $b$ are some real nonnegative numbers, and for these numbers
the inequalities
\begin{eqnarray}    
&&\left(p_1(a)-{1}/{2}\right)^2+\left(p_2(a)-{1}/{2}\right)^2
  +\left(p_3(a)-{1}/{2}\right)^2\leq {1}/{4},\nonumber\\[-2mm]
  &&\label{eq.4}\\[-2mm]
&&\left(p_1(b)-{1}/{2}\right)^2+\left(p_2(b)-{1}/{2}\right)^2
  +\left(p_3(b)-{1}/{2}\right)^2\leq {1}/{4}\nonumber
\end{eqnarray}{
hold. To show this, we construct the following map of an arbitrary
Hermitian matrix $H=H^\dagger$ onto a nonnegative Hermitian matrix
$\rho(x)$ with unit trace. The matrix elements of the matrix
$\rho(x)$ depend on the parameters $-\infty\leq x\leq
\infty$.

Now we express the matrix $\rho(x)$ in terms of matrix elements of
the matrix $H$ as follows:
\begin{eqnarray}    
&&\rho_{11}(x)=\frac{H_{11}+x}{H_{11}+H_{22}+2x}\,,\qquad
\rho_{12}(x)=
  \frac{H_{12}}{H_{11}+H_{22}+2x}\,,\nonumber\\[-2mm]
&&\label{eq.5}\\[-2mm]
&&\rho_{21}(x)=\frac{H_{21}}{H_{11}+H_{22}+2x}\,,\qquad\rho_{22}(x)=
  \frac{H_{22}+x}{H_{11}+H_{22}+2x}\,.\nonumber
\end{eqnarray}
It is obvious that for $x\geq|x_0|$, where $x_0$ is the smallest of
two eigenvalues of the Hermitian matrix $H$, the matrix $\rho(x)$
satisfies the conditions $\rho(x)=\rho^{\dagger}(x)$,
$\mbox{Tr}\rho(x)=1$, and $\rho(x)\geq 0$. This means that the matrix
$\rho(x)$ for such values of the parameter $x$ can be interpreted as
the density matrix of the qubit state and, in view of this fact, it
can be presented in the form (\ref{A}).

One can check that, if one takes two different values of the
parameter $x$, e.g., $x=a$ and $x=b$, where $a,b\geq|x_0|$, the
matrix of observable $H$ can be expressed in terms of the matrix
elements of the two density matrices $\rho(a)$ and $\rho(b)$,
namely,
\begin{eqnarray}
&&H_{11}=\frac{a p_3(b)\left(1-2p_3(a)\right)-b
p_3(a)\left(1-2p_3(b)\right)}{p_3(a)-p_3(b)}\,,\nonumber\\
&&  H_{11}+H_{22}=\frac{a-b+2\left(b p_3(b)-a
  p_3(a)\right)}{p_3(a)-p_3(b)}\,,\nonumber\\
  &&\label{eq.6}\\[-2mm]
&&H_{12}=(H_{11}+H_{22}+2a)\rho_{12}(a)=(H_{11}+H_{22}+2b)\rho_{12}(b)\,,\nonumber\\
&&  H_{21}=H_{12}^{\ast}\,.\nonumber
\end{eqnarray}
These relations provide the matrix elements of the observable $H$ in
terms of the probabilities $p_1(a)$, $p_2(a)$, $p_3(a)$, $p_1(b)$,
$p_2(b)$, and $p_3(b)$. The observable $H$ can be, e.g., the
Hamiltonian; also
\begin{equation} \label{eq.7}
\rho_{12}(a)=p_1(a)-ip_2(a)-({1-i})/{2}.
\end{equation}
If $~H=\sigma_z=\left(\begin{array}{cc}
  1&0\\
  0&-1\end{array}\right),~$ then
$~
\rho(x)=\left(\begin{array}{cc}
  2^{-1}+(2x)^{-1}&0\\
  0&2^{-1}-(2x)^{-1}\end{array}\right)~$ at  $~x> 1.$

\section{Spin-1/2 Tomography}  
The tomographic probability distribution $w(m,\vec n)$ describing
the qubit state was defined in~\cite{DodPLA,OlgaJetp} as diagonal
matrix elements of the density matrix in the rotated reference
frame, i.e.,
\begin{equation} \label{eqA1}
w(m,\vec n)=\langle m|u\rho u^{\dagger}|m\rangle,
\end{equation}
where the unitary matrix $u_{j k}$ $(j,k=1,2)$ is expressed in terms
of the Euler angles as follows:
\begin{equation} \label{eq.A2}
u_{jk}=\left(\begin{array}{cc}
  \cos\frac{\theta}{2}e^{i(\phi+\psi)/2}&\sin\frac{\theta}{2}e^{i(\phi-\psi)/2}\\
  -\sin\frac{\theta}{2}e^{-i(\phi-\psi)/2}&\cos\frac{\theta}{2}e^{-i(\phi+\psi)/2}\\
\end{array}\right),
\end{equation}
and $m=\pm1/2$ is the spin projection on the direction determined by
the unit vector $\vec n$,
$$\vec n=(\sin\theta\cos\phi,\sin\theta\sin\phi,\cos\theta).$$

The function $w(m,\vec n)$, called the tomographic probability
distribution, gives the probability to have the spin-projection $m$
on the direction $\vec n$. The density matrix $\rho$ is expressed in
terms of the tomographic probability distribution $w(m,\vec
n)$~\cite{DodPLA,OlgaJetp}. Since there are only three parameters
determining the density matrix $\rho$, information contained in the
qubit tomogram $w(m,\vec n)$, where $\vec n_1=(1,0,0)$, $\vec
n_2=(0,1,0)$, and $\vec n_3=(0,0,1)$, is sufficient to obtain the
density matrix.

This matrix is given by (\ref{A}), where $p_1$, $p_2$, and $p_3$ are
the probabilities to have $m=+1/2$ on the above directions. An
arbitrary tomogram is expressed in terms of the probabilities $p_1$,
$p_2$, and $p_3$ ($p=p_1+i p_2$, $\gamma=(1+i)/{2}$)  as follows:
$$
w(m,\vec n)=\left(\left(\begin{array}{cc}
  u_{11}&u_{12}\\
  u_{21}&u_{22}\\
\end{array}\right)
\left(\begin{array}{cc}
  p_3&p^\ast-\gamma^\ast\\
  p-\gamma&1-p_3\\
\end{array}\right)
\left(\begin{array}{cc}
  u_{11}^\ast & u_{21}^\ast\\
  u_{12}^\ast & u_{22}^\ast\\
\end{array}\right)\right)_{m m}.
$$
Thus, the tomogram (\ref{eqA1}) reads
\begin{equation} \label{8'}
w(+1/2,\vec n)=(\vec p-\vec p_0)\vec n+1/2,
\end{equation}
where $\vec p_0=\left(1/2,1/2,1/2\right)$ and $w(-1/2,\vec
n)=1-w(+1/2,\vec n)$. Thus, we describe an arbitrary qubit state by
means of three probabilities in any rotated reference frame.

Applying the formula obtained to the spin-$1/2$ tomogram, we can
express the density matrix $\rho_u=u\rho u^\dagger$ in terms of
probabilities $p_1$, $p_2$, and $p_3$ written as components of the
vector $\vec p=(p_1,p_2,p_3)$ and vectors $\vec {x'}$, $\vec {y'}$,
and $\vec {z'}$ obtained by rotations of three basis vector  $\vec
x$, $\vec y$, and $\vec z$ given by the unitary matrix $u$. This
means that we have three orthogonal vectors $\vec {x'}$, $\vec
{y'}$, and $\vec {z'}$ obtained by the rotation $O(3)$ from the
initial vectors $\vec x$, $\vec y$, and $\vec z$, i.e., $\vec
{x'}=O\vec x$, $\vec {y'}=O\vec y$, and $\vec {z'}=O\vec z$. Using
these expressions, one can get the probabilities $p_1'$, $p_2'$, and
$p_3'$ determined by the matrix $\rho'=u\rho u^\dagger$ in the form
$$
p_1'=L_{11}p_1+L_{12}p_2+L_{13}p_3+C_1,$$
$$p_2'=L_{21}p_1+L_{22}p_2+L_{23}p_3+C_2,$$
$$p_3'=L_{31}p_1+L_{32}p_2+L_{33}p_3+C_3,
$$
where the matrix $L$ and vector $\vec C$ are expressed in terms of
the unitary matrix as follows:
\begin{eqnarray*}
&&L_{31}=u_{12}u_{11}^\ast+u_{11}u_{12}^\ast,\quad
L_{32}=i\left(u_{12}u_{11}^\ast-u_{11}u_{12}^\ast\right),\quad
L_{33}=|u_{11}|^2-|u_{12}|^2,\\
&&L_{13}=\mbox{Re}\left(u_{11}u_{21}^\ast\right)-\mbox{Re}\left(u_{12}u_{22}^\ast\right),\quad
L_{12}=\mbox{Re}\left(i u_{12}u_{21}^\ast\right)-\mbox{Re}\left(iu_{11}u_{22}^\ast\right),\\
&&L_{11}=\mbox{Re}\left(u_{12}u_{21}^\ast\right)+\mbox{Re}\left(u_{11}u_{22}^\ast\right),
\quad
L_{23}=\mbox{Im}\left(u_{12}u_{22}^\ast\right)-\mbox{Im}\left(u_{11}u_{21}^\ast\right),\\
&&L_{22}=\mbox{Im}\left(i u_{11}u_{22}^\ast\right)-\mbox{Im}\left(i
u_{12}u_{21}^\ast\right),\quad
L_{21}=-\mbox{Im}\left(u_{12}u_{21}^\ast\right)-\mbox{Im}\left(u_{11}u_{22}^\ast\right),\\
&& C_1=\mbox{Re}\left(-\gamma u_{12}u_{21}^\ast-\gamma^\ast
u_{11}u_{22}^\ast+u_{12}u_{22}^\ast+\gamma\ast\right),\\
&&C_2=\mbox{Im}\left(\gamma u_{12}u_{21}^\ast+\gamma^\ast
u_{11}u_{22}^\ast-u_{12}u_{22}^\ast-\gamma\ast\right),\\
&& C_3=-\gamma u_{12}u_{11}^\ast-\gamma^\ast
u_{11}u_{12}^\ast+|u_{12}|^2.
\end{eqnarray*}
The density matrix $\rho'$ is expressed in terms of probabilities
$(p_1,p_2,p_3)=\vec p$ as
$$
\rho'=(\sigma_0/2)+(\vec p-\vec p_0)\left(\vec {z'}\sigma_z+
  \vec {x'}\sigma_x+\vec {y'}\sigma_y\right)
$$
with the vector $\vec p_0=(1/2,1/2,1/2)$. This form shows that the
probabilities $p_1'=(\vec p-\vec p_0)\vec {x'}$, $p_2'=(\vec p-\vec
p_0)\vec {y'}$, and $p_3'=(\vec p-\vec p_0)\vec {z'}+1/2$ are the
probabilities to obtain the spin projection $m=+1/2$ along the
directions given by vectors $\vec {x'}$, $\vec {y'}$, and $\vec
{z'}$, respectively.

For unitary transform of the density matrix
$~\rho\longrightarrow\rho'=\sum_s {\cal P}_s u_s\rho u_s^\dagger, $
where we have the probability distribution $1\geq{\cal P}_s\geq0$,
$\sum_s{\cal P}_s=1$, and $u_s$ are unitary matrices, the tomogram
$w(+{1}/{2},\vec n)$ converts into
$$
w\big(+{1}/{2},\vec n\big)=\sum_s\sum_{j,k=1}^3{\cal P}_s(\vec
p-\vec p_0)_j O^{(s)}_{jk} n_k +{1}/{2}.
$$
Here, $O_{j k}^{(s)}$ are real orthogonal 3$\times$3 matrices
$(O^T=O^{-1})$. Then the new density matrix $\rho'$ reads
$$
\rho'=\frac{1}{2}\sigma_0+(\vec p-\vec p_0)\sum_s\left[\left(\big(\sigma_z O^{(s)}{\cal P}_s\big)
  \vec z\right)\sigma_z+\big(\sigma_x{\cal P}_s O^{(s)}\big)\vec x+
  \big(\sigma_y{\cal P}_s O^{(s)}\big)\vec y\right].
$$
The 3$\times$3 matrix $\sum_s{\cal P}_s O^{(s)}$ is the convex sum
of orthogonal 3$\times$3 matrices $O^{(s)}$. In the above formula,
it acts on the basis vectors $\vec x$, $\vec y$, and $\vec z$.

Thus, the new probabilities obtained due to the unitary transform
are linear combinations of the $\vec p$ components, i.e.,
\begin{eqnarray*}
&&p_1'=\sum_s\sum_{k=1}^3{\cal P}_s O_{k_1}^s\left(\vec p-\vec
p_0\right)_k\,, \\
&& p_2'=\sum_s\sum_{k=1}^3{\cal P}_s
O_{k_2}^s\left(\vec p-\vec p_0\right)_k\,, \\
&&p_3'=\sum_s\sum_{k=1}^3{\cal P}_s O_{k_3}^s\left(\vec p-\vec
p_0\right)_k+1/2. \end{eqnarray*}
The expressions obtained describe the quantum channels for qubit
states in the probability representation.

\section{Spin-1/2 Observable Tomograms}    
To provide the probability description of spin observables, we
construct a tomogram of the matrix $\rho(x)$ for an arbitrary
parameter $x$. First, we introduce the function
\begin{equation} \label{K1}
w(m,\vec n,x)=\left[\left(\begin{array}{cc}
  u_{11}&u_{12}\\
  u_{21}&u_{22}\\
  \end{array}\right)
\left(\begin{array}{cc}
  \rho_{11}(x) &\rho_{12}(x)\\
  \rho_{21}(x) &\rho_{22}(x)\\
  \end{array}\right)
\left(\begin{array}{cc}
  u_{11}^\ast &u_{12}^\ast\\
  u_{21}^\ast &u_{22}^\ast\\
  \end{array}\right)
\right]_{m m}, \quad m=\pm1/2.
\end{equation}
Here, we use the map of the matrix indices
$$1\,1\leftrightarrow 1/2\,\,1/2,\quad
1\,2\leftrightarrow1/2\,\,-1/2,\quad
2\,1\leftrightarrow-1/2\,\,1/2,\quad
2\,2\leftrightarrow-1/2\,\,-1/2.$$ For $x>|x_o|$, the function
$w(m,\vec n,x)\geq 0$ satisfies the normalization condition
$\sum_{m=-1/2}^{1/2}w (m,\vec n, x)=1.$

Following the derivation of probabilities for the given density
matrices, we introduce the probabilities $P_3(a)$ and $P_3(b)$ given
by Eq.~(\ref{eq.4}) as follows:
\begin{equation} \label{eq.k3}
P_3(a)=\frac{H_{11}+a}{H_{11}+H_{22}+2a}\,,\qquad
P_3(b)=\frac{H_{11}+b}{H_{11}+H_{22}+2a}\,.
\end{equation}
Also the probabilities $P_1(a)$, $P_2(a)$, $P_1(b)$, and $P_2(b)$
are determined by the relations
\begin{equation} \label{eq.k4}
P_1(a)-i P_2(a)-\gamma^\ast=\frac{H_{12}}{H_{11}+H_{22}+2a}\,,\qquad
  P_1(b)-iP_2(b) -\gamma^\ast=\frac{H_{12}}{H_{11}+H_{22}+2b}\,.
\end{equation}
Thus, we obtain the tomograms of observable $H$ for $\vec n=0,0,1$;
they read
\begin{equation} \label{k4}
w(+{1}/{2},\vec
n,a)=P_3(a)=\frac{H_{11}+a}{H_{11}+H_{22}+2a}\,,\quad
  w(+{1}/{2},\vec n,b)=P_3(b)=\frac{H_{11}+b}{H_{11}+H_{22}+2b}\,.
\end{equation}
Also for an arbitrary $\vec n$, we have
$$
w(+{1}/{2},\vec n,a)=\left(\vec P(a)-\vec P_0\right)\vec n +{1}/{2},
\qquad
  w(+{1}/{2},\vec n,b)=\left(\vec P(b)-\vec P_0\right)\vec n +{1}/{2}.
$$

\section{Triangle Geometry of Tomographic Probabilities of Observables
in the Quantum Suprematism Picture}  
 Since we introduce the map of the spin-1/2 observable onto two density matrices $\rho(a)$ and 
$\rho(b)$, we can apply the tomographic description of the density
matrices and known geometrical properties of qubit states formulated
in terms of the triangle geometry using the {\it Triada of
Malevich's squares} \cite{Malevich} 
in the quantum suprematism
picture. The specific feature of the probability representation of
the qubit observable is related to the fact that the Hermitian
matrix $H$ is connected with two density matrices; this means that
we use the probabilities $P_1(a)$, $P_2(a)$, $P_1(b)$, $P_2(b)$, and
$P_3(a)$, $P_3(b)$ to describe the observable.

Thus, the probabilities are associated with vertices $A_1(a)$,
$A_2(a)$, $A_3(a)$ and $A_1(b)$, $A_2(b)$, $A_3(b)$ of the triangles
shown in Figs.~1 and 2. These vertices are located on the sides of
the equilateral triangle with a side length of $\sqrt
2$ \cite{Malevich}. 
The two triangles of Malevich's squares
determined by the probabilities  $0\leq P_1(a)$, $P_2(a)$, $P_3(a)$,
$P_1(b)$, $P_2(b)$, $P_3(b)\leq 1$ are shown in Figs.~3 and 4.

\begin{figure}
  \centering
  \includegraphics[width=90mm]{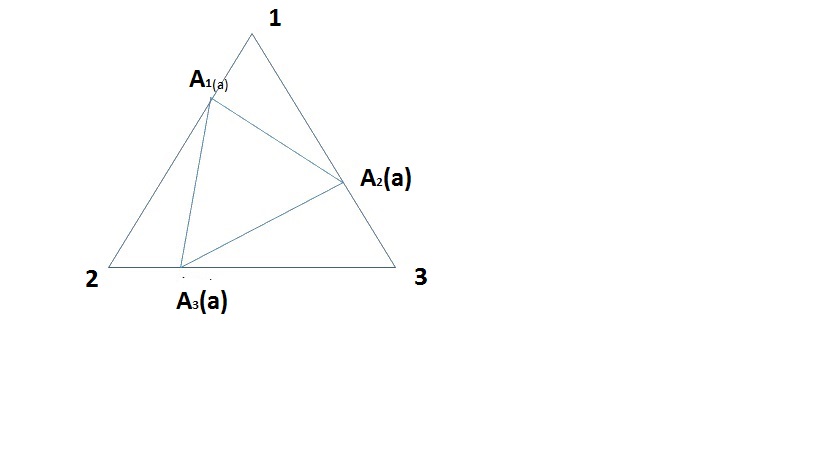}\\
  \caption{Triangle $A_1(a)A_2(a)A_3(a)$ corresponding to three probabilities 
$P_1(a)$, $P_2(a)$, and $P_3(a)$ determining the density matrix
$\rho(x=a)$.}\label{1}
\end{figure}

\begin{figure}
  \centering
  \includegraphics[width=90mm]{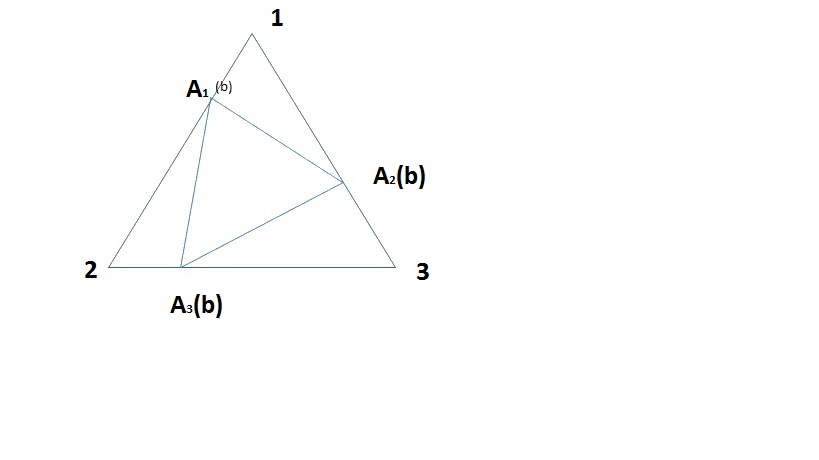}\\
  \caption{ Triangle $A_1(b)A_2(b)A_3(b)$ corresponding to three probabilities 
$P_1(b)$, $P_2(b)$, and $P_3(b)$ determining the density matrix
$\rho(x=b)$.}\label{3}
\end{figure}

\begin{figure}
  \centering
  \includegraphics[width=90mm]{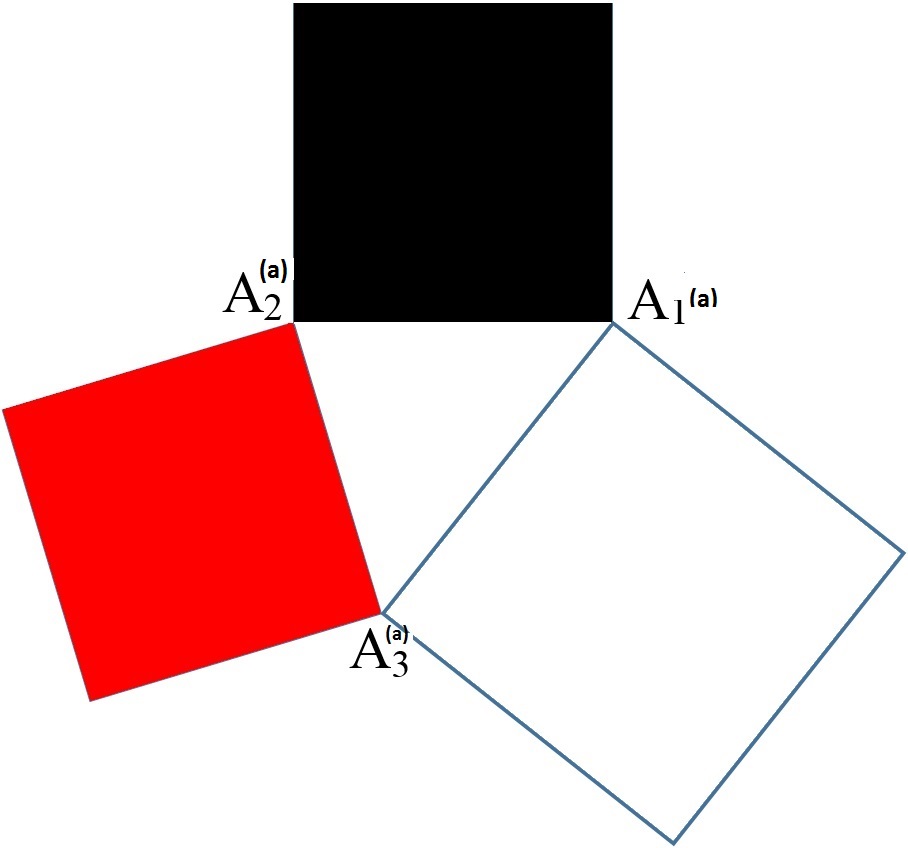}\\
  \caption{Triada of Malevich's squares containing complete information on the
density matrix $\rho(x=a)$.}\label{3}
\end{figure}

\begin{figure}
  \centering
  \includegraphics[width=90mm]{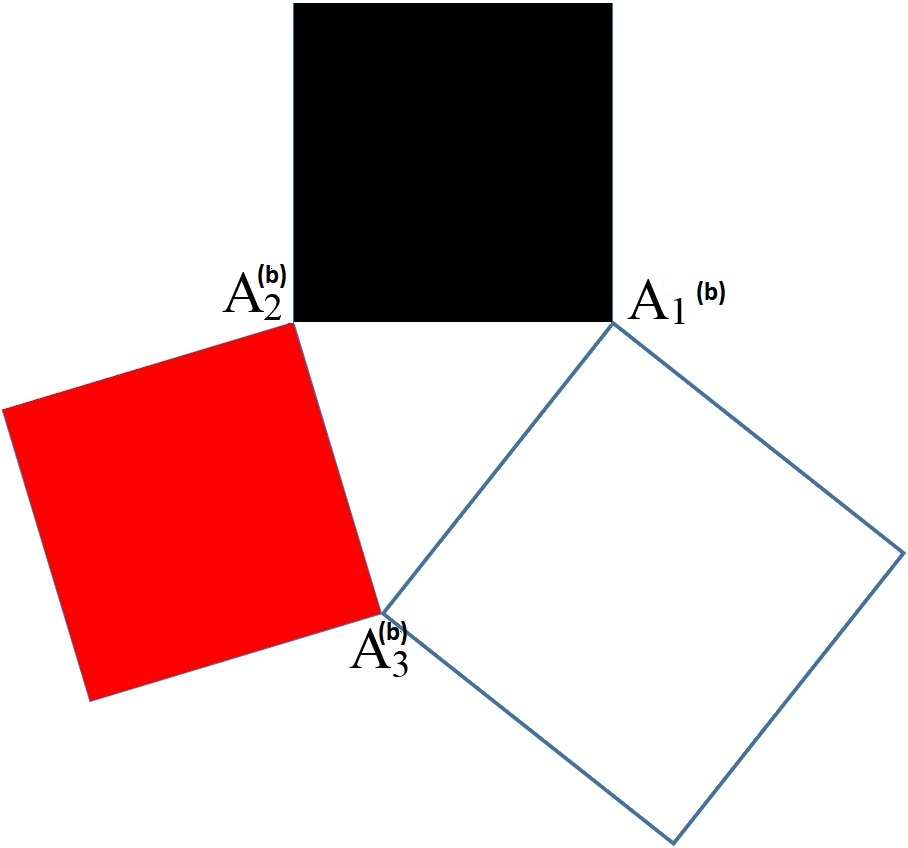}\\
  \caption{ Triada of Malevich's squares containing complete information on the 
density matrix $\rho(x=b)$.}\label{4}
\end{figure}

\begin{figure}
  \centering
  \includegraphics[width=120mm]{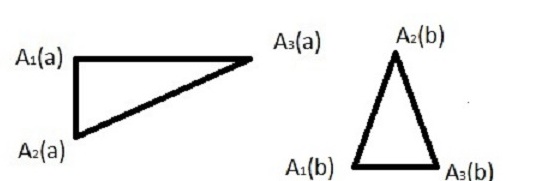}\\
\caption{Two triangles corresponding to the spin projection of observable
$\sigma _x$ with vertices $A_1(a)$, $A_2(a)$, $A_3(a)$~(left) and
$A_1(b)$, $A_2(b)$, $A_3(b)$~(right) in triangle geometrical picture
of quantum observables. }\label{5}
\end{figure}

The sums of areas of the squares read
\begin{eqnarray*}    
&&S_a=2\big[3\big(1-p_1(a)-p_2(a)-p_3(a)\big)+2p_1^2(a)+2p_2^2(a)+2p_3^2(a)\big.\\
&&\big.+p_1(a)p_2(a)+p_2(a)p_3(a)+p_3(a)p_1(a)\big], \\
&&S_b=2\big[3\big(1-p_1(b)-p_2(b)-p_3(b)\big)+2p_1^2(b)+2p_2^2(b)+2p_3^2(b)\big.\\
&&+p_1(b)p_2(b)+p_2(b)p_3(b)+p_3(b)p_1(b)\big].
\end{eqnarray*}
There are bounds for the sums of the areas due to hidden quantum
correlations in artificial qubit states $\rho(a)$ and $\rho(b)$
associated with the observable $H$. 
The minimum area is $3/2$. If the observable $H$ is the Pauli matrix
$\sigma_z$, i.e., it corresponds to the spin projection on the
$z$~axis, $~H=\sigma_z=\left(\begin{array}{cc}
  1 &0\\
  0 &-1\\
\end{array}\right),
$ the triangles with vertices $A_1(a)$, $A_2(a)$, $A_3(a)$ and
$A_1(b)$, $A_2(b)$, $A_3(b)$ look as shown in Fig.~5. The length of
sides $A_1(a)A_2(a)$ and $A_1(b)A_2(b)$ is equal to $\sqrt2/2$. In
this case, the maximum area $S(a)$ and $S(b)$ is $5/2$, and the
minimum area is $3/2$.

The area $S(a)=3/2$ corresponds to the matrix
$~\rho(a)=\left(\begin{array}{cc}
  1/2 &0\\
  0 &1/2\\
\end{array}\right).
$

\section{The Evolution Equation for Qubit Observables in the Probability
Representation}  
 Given an observable $A_{jk}$ in the matrix
form and a Hamiltonian $H_{jk}$ $(j,k=1,2)$, the Heisenberg equation
for the observable $A(t)$ reads
\begin{equation} \label{Ev1}
\frac{\partial A(t)}{\partial t}=i[H,A(t)].
\end{equation}
The corresponding ``density matrix'' $\rho(x,t)$
\begin{equation} \label{Ev2}
\rho(x,t)=\frac{1}{\mbox{Tr}\,A(t)+2x}(A(t)+x1_2)
\end{equation}
satisfies the evolution equation
\begin{equation} \label{Ev3}
\frac{\partial\rho(x,t)}{\partial t}=i[H,\rho(x,t)].
\end{equation}
To obtain this equality, we employed the property
$
\frac{\partial}{\partial t}(\mbox{Tr}\, A(t)+2x)=0,$
which follows from Eq.~(\ref{Ev1}). Since $\rho(x,t)$ can be
expressed in terms of probabilities $p_1(x,t)$, $p_2(x,t)$, and
$p_3(x,t)$,
\begin{equation} \label{Ev5}
\rho(x,t)=\left(\begin{array}{cc}
  p_3(x,t)&p_1(x,t)-i p_2(x,t)-\gamma^\ast\\
  p_1(x,t)+i p_2(x,t)-\gamma & 1-p_3(x,t)\\
\end{array}\right),
\end{equation}
the evolution equation (\ref{Ev3}) can be presented as the system of
equations for the probability vector
$~\vec p(x,t)=\left(\begin{array}{ccc}
  p_1(x,t)\\
  p_2(x,t)\\
  p_3(x,t)\\
\end{array}\right)$
given by the following expression:
\begin{equation} \label{Ev7}
\frac{d \vec p(x,t)}{d t}=L \vec p(x,t)+\vec C.
\end{equation}
Here, the 3$\times$3 matrix $~L~$ reads $~L=\left(\begin{array}{ccc}
  0\,&\,H_{11}-H_{22}\,&\,-2\,\mbox{Im}\,H_{21}\\
  H_{22}-H_{11}\,&\,0\,&\,2\,\mbox{Re}\,H_{21}\\
  2\,\mbox{Im}\,H_{21}\,&\,-2\,\mbox{Re}\,H_{21}\,&\,0\\
\end{array}\right),$
the three-vector $~\vec C~$ is $~
\vec C=\left(\begin{array}{ccc}
  \mbox{Im}\,H_{21}+(H_{22}-H_{11})/2\\
  -\mbox{Re}\,H_{21}+(H_{11}-H_{22})/2\\
  2\,\mbox{Im}\,(\gamma H_{12})\\
\end{array}\right),
$
and the number $H_{j k}$ can be expressed in terms of probabilities
$p_1(a)$, $p_2(a)$, $p_3(a)$ and $p_1(b)$, $p_2(b)$, $p_3(b)$ given
by (\ref{eq.k3})--(\ref{k4}). This means that the equation for
$\rho(x,t)$ has the form containing only the probabilities
$p_1(x,t)$, $p_2(x,t)$, $p_3(x,t)$, $p_1(a)$, $p_2(a)$, $p_3(a)$,
and $p_1(b)$, $p_2(b)$, $p_3(b)$.

The unitary evolution preserves the eigenvalues of observable $A$.
Due to this, the vector $\vec p(x,t)$ for chosen parameters $a$ and
$b$ has the nonnegative components varying in the domain $0\leq
p_{j}(x,t)\leq1$ and satisfying the inequalities providing the
nonnegativity condition for the density matrix $\rho(x,t)$.

\section{Conclusions}  
To conclude, we list the main results of this work.

We presented qubit states by the density matrix with matrix elements
expressed in terms of three measurable probabilities of spin-$1/2$
projections in three perpendicular directions and obtained compact
formula~(\ref{8'}) for spin tomogram; the matrix of an arbitrary
observable is expressed in terms of two probability distributions
given by (\ref{eq.6}) and (\ref{eq.7}). Also we demonstrate the
Heisenberg evolution equation for an arbitrary observable as a
system of linear kinetic evolution equations for probabilities given
by Eq.~(\ref{Ev7}); the coefficients of the equations are also
expressed in terms of the probabilities describing arbitrary
Hamiltonian matrix elements. Thus, we formulated all the ingredients
of quantum mechanics
--- states, observables, quantum evolution equation --- in terms of
probabilities; with the states and observables being identified in a
conventional formulation of quantum mechanics with vectors in
Hilbert spaces and operators acting in the Hilbert spaces. Note that
the probabilities in classical probability theory are related to the
other geometrical structure -- simplexes.

We found the possibility to map the quantum states and observables
formulated using the Hilbert space characteristics onto the
probabilities associated with characteristics of the simplexes. This
is done for the spin-$1/2$ system but we conjecture that the
construction of an analogous map can be found for arbitrary quantum
systems. We present such construction for qutrits in a future
publication.

\section*{Acknowledgments} 
The formulation of the problem of the evolution equation for qubit
observables in the probability representation  and the results of
Sec.~2 are due to V. I. Man'ko, who is supported by the Russian
Science Foundation under Project No.~16-11-00084; this work was
partially performed at the Moscow Institute of Physics and
Technology. The authors thank A.~Avanesov for correcting some
formulas of~\cite{Malevich}.

\end{document}